\documentclass[sigconf]{acmart}
\usepackage{natbib}
\usepackage{graphicx}
\usepackage{caption}
\usepackage{subcaption}
\usepackage{multirow}
\usepackage{hyperref}
\usepackage[]{algorithm2e}
\usepackage{cite}
\usepackage{booktabs}			

\usepackage{latexsym}			
\usepackage{fancyvrb}			

\usepackage{amsmath}

\usepackage{stmaryrd}			

\usepackage{subfiles}

\usepackage{sty/term}				
\usepackage{sty/nospace}			
\usepackage{sty/prolog}				
\usepackage{sty/local}				

\usepackage{graphicx}			

\usepackage{color}	
\usepackage{booktabs}
\usepackage{multirow}
\usepackage{cite}
\usepackage{framed}
\usepackage{textpos}
\usepackage{float}
\usepackage{amsbsy}
\usepackage{float}
\usepackage{array}

\newif\ifdoublespace\doublespacefalse 

\begin{document}
%

\title{Answer Graph: Factorization Matters in Large Graphs}

\author{Zahid Abul-Basher}
\affiliation{%
	\institution{University of Toronto}
	\city{Toronto}
	\country{Canada}
}
\email{zahid@mie.utoronto.ca}

 \author{Nikolay Yakovets}
 \affiliation{%
   \institution{Eindhoven University of Technology}
   \city{Eindhoven}
   \country{The Netherlands}
 }
 \email{hush@tue.nl}

 \author{Parke Godfrey}
 \affiliation{%
   \institution{York University}
   \city{Toronto}
   \country{Canada}
 }
 \email{godfrey@yorku.ca}

 \author{Stanley Clark}
 \affiliation{%
   \institution{Eindhoven University of Technology}
   \city{Eindhoven}
   \country{The Netherlands}
 }
 \email{s.clark@tue.nl}

 \author{Mark Chignell}
 \affiliation{%
   \institution{University of Toronto}
   \city{Toronto}
   \country{Canada}
 }
 \email{chignell@mie.utoronto.ca}

\begin{abstract}
Our \emph{answer-graph method}
to evaluate \sparql\ \emph{conjunctive queries} (\cq s)
finds a \emph{factorized} answer set first, an \emph{answer graph},
and then finds the embedding tuples from this.
This approach can reduce greatly the cost to evaluate \cq s.
This affords a second advantage:
we can construct a cost-based planner.
We present the answer-graph approach,
and
overview our prototype system, \wireframe.
We then offer \emph{proof of concept}
via a micro-benchmark
over the YAGO2s dataset
with two prevalent \emph{shapes} of queries,
snowflake and diamond.
We compare \wireframe's performance over these
against
    \postgresql,
    \virtuoso,
    \monetdb,
        and
    \neofourj\
to illustrate the performance advantages
of our answer-graph approach.

\end{abstract}


\maketitle



\section{Introduction}
\begin{figure}[t]
	\center
	\includegraphics[width=\linewidth]{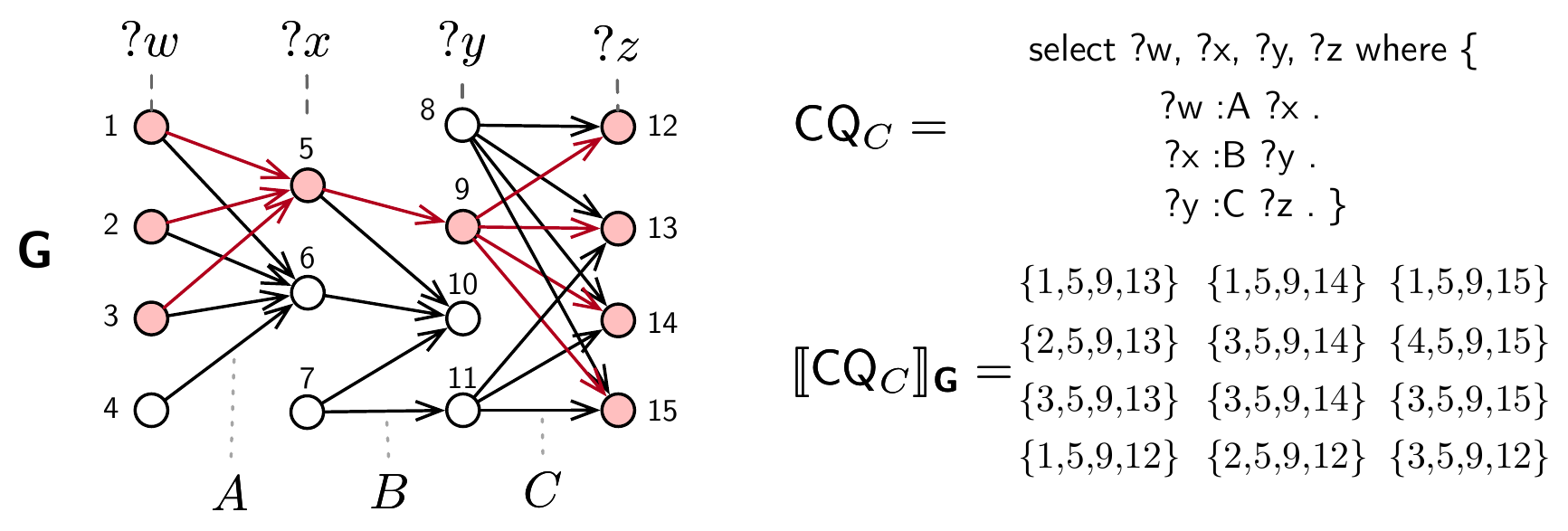}
	\caption{Example of an answer graph (shaded red).}
	\label{fig:factorization_example}
\end{figure}
\begin{figure*}[t]
	\center
	\includegraphics[width=\textwidth]{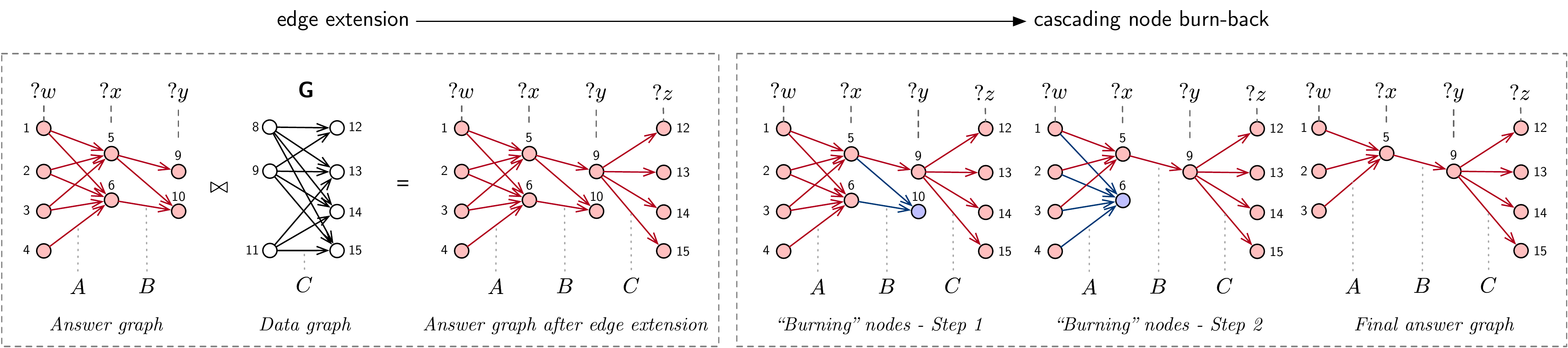}
	\caption{An example of the evaluation model
	for the answer graph generation.}
	\label{fig:evaluation_model_example}
\end{figure*}
Science is, of course, driven by observation.
It is now also becoming ever more data driven.
Some of the datasets involved are unimaginably large.
The data is often wildly heterogeneous,
and rarely well structured as in business applications.
This demands new skills, methods, and approaches of scientists,
and challenges computer scientists with devising
new data models, query languages, systems, and tools
that better support this.

Graph-like data has become prevalent
among scientific data stores and elsewhere.
The data-science research community has begun to focus
on how best to support the management of graph data
and its analysis.
One \emph{data model} for graph databases
is the \emph{Resource Description Framework} (\rdf)
\citep{W3C},
paired with the \emph{query language} \sparql\ 
\citep{sparql}.
These have evolved as W3C standards,
initially for addressing the Semantic Web.
An \rdf\ store conceptually consists of a set of \textit{triples}
to represent a directed, edge-labeled multi-graph.
The triple \tuple{\var{s}, \var{p}, \var{o}}\ represents
the directed, labeled edge
from \emph{subject} node ``\var{s}''
to \emph{target} node ``\var{o}''
with \emph{label} (\emph{predicate}) ``\var{p}''.
In \rdf,
nodes have unique identity.
The semantics, however, is carried by the labels
and how the nodes are connected.
%
The \uniprot\
\citep{UniProt}
\sparql\ \textsc{Endpoint} (dataset)
\citep{UniProtEndpoint},
for example,
consists of 63,376,853,475 \rdf\ triples
as of this writing.
\uniprot\ (\underline{Uni}versal \underline{Prot}ein resource)
is a freely accessible, popular repository of protein data.%

The \sparql\ query language provides a formal way to query
over such graph databases.
Types of \sparql\ queries can be thought about
as small graphs themselves,
so-called \emph{query graphs}.
In a \sparql\ \emph{conjunctive query} (\cq),
the ``nodes'' are the query's \emph{binding variables}
and the ``edges'' between these are the labels to be matched.
An ``answer'' to a \cq\ on a \emph{data graph} \dg\
(that is, the graph database),
denoted as \eval{\cq}{\dg},
is a homomorphic \emph{embedding} of the query graph
into the data graph
that matches the query's labels
to the data graph's labeled edges.
An answer is
then a tuple of node ID's
as a binding of the query's node variables.
As such,
each answer can be considered as a sub-graph matched
in the data graph.

A \cq\ can be quite expensive to evaluate
and require extreme resources,
given both the potentially immense size of the data graph
and the relative complexity of the \cq's query graph.
The challenge is to reduce the expense and needed machinery.
We present a novel approach to \emph{query optimization}
and \emph{evaluation} for \cq s
that consists of two parts:
1) \emph{factorization};
and
2) \emph{cost-based plan enumeration} via dynamic programming.
The answer set of a \cq\ is a set of embeddings.
This in itself is \emph{not} a graph.
Instead, as an intermediate step,
we do find an \emph{answer graph},
the subset of edges from the data graph
that suffices to compose the \cq's embeddings.

A number of \rdf\ systems have been developed 
with quite different architectures.
These include
    triple stores (\emph{e.g.}, RDF-3X \citep{neumann:2008:rdf}), 
    property tables (\emph{e.g.}, BitMat \citep{Atre:2010:bitmat}), 
    column-based stores (\emph{e.g.}, \citep{Abadi:2007:SSW}),
        and 
    graph-based stores (\emph{e.g.}, gStore \citep{Zou:2011:gstore}). 
Our answer-graph approach is generic,
and can be implemented
within any of these architectures. We posed this
idea initially in \citep{godfrey:2017:wireframe}. We develop it here. 
We demonstrate its key advantages
via our prototype system, \wireframe,
and a micro-benchmark.
\section{The Answer-graph Approach}

In \citep{bakibayev:2012:fdb},
the authors introduce the concept of \emph{factorization}
as a query-optimization technique for relational databases.
Their technique is designed, and works exceptionally well,
for schema and queries
for which cross products of projections of the answer tuples
all show up as answer tuples.
This happens, for instance, in schema not in \emph{fourth normal form}.
Evaluating for these projected tuples first
and then cross-producting them later can be
a much more efficient strategy.
Deciding how best to \emph{factorize}%
---%
how to project into sub-tuples%
---%
is difficult,
however.

For \cq s,
this last part is trivial:
the factorization of the embedding tuples
is fully down to component node pairs,
corresponding to the labeled edges.
This is our \emph{answer graph}.%
\footnote{%
	This is demonstrably true 
	when the \cq\ is \emph{tree} shaped.
	This is arguable when the \cq\ has cycles.
	In the latter case,
	the factorization can be characterized as projections
	to tuples of node pairs and node triples (\emph{triangles}).
}
While factorization is sometimes a significant win
for evaluating relational queries,
it is virtually always a win for evaluating graph \cq s.

An answer graph, \ag,
for a \cq\
is a subset of the data graph \dg\
that suffices to compute the embeddings for the \cq.
We call the \emph{minimum} such subset
the \emph{ideal answer graph}, \iag.
The \iag\ is often quite small,
significantly smaller than the set of embeddings,
and extremely much smaller than \dg.
Thus,
evaluating a \cq's embeddings
in two steps%
---%
first, find its \iag,
then compose the embeddings
(which we call \emph{defactorization})
from the \iag\ rather than from \dg%
---%
can be significantly more efficient.


Consider the data graph \dg\ and the \emph{chain} query \cq[C]\
in Fig.~\ref{fig:factorization_example}.
\cq[C]\ which finds all node-tuples
\tuple{\var{w}, \var{x}, \var{y}, \var{z}}\
from \dg\ such that
\tuple{\var{w}, \var{x}}
is connected by an edge labeled \var{A},
\tuple{\var{x}, \var{y}}
by \var{B},
and
\tuple{\var{y}, \var{z}}
by \var{C}.
Due to multiplicity
from \var{A}-edges \emph{fanning in} to,
and \var{C}-edges \emph{fanning out} of,
\var{B}\ values,
the embedding set is \emph{twelve} tuples.
Meanwhile,
our answer graph consists of \emph{eight} labeled node pairs (shown in red).
Such differences are greatly magnified when on a larger scale.

Our answer-graph approach affords us a second key advantage.
We can devise a cost-based query optimizer
based on dynamic programming to construct a \emph{query plan}.
A plan for us is simply a specified order of the \cq's query edges
in which to evaluate to matching answer-graph edges.
Our evaluation strategy for such plans is explained next,
and our \wireframe\ optimizer for choosing plans
in Section \ref{sec:framework}.
\section{The Evaluation Model}

\begin{figure*}[t]
	\center
	\includegraphics[width=\textwidth]{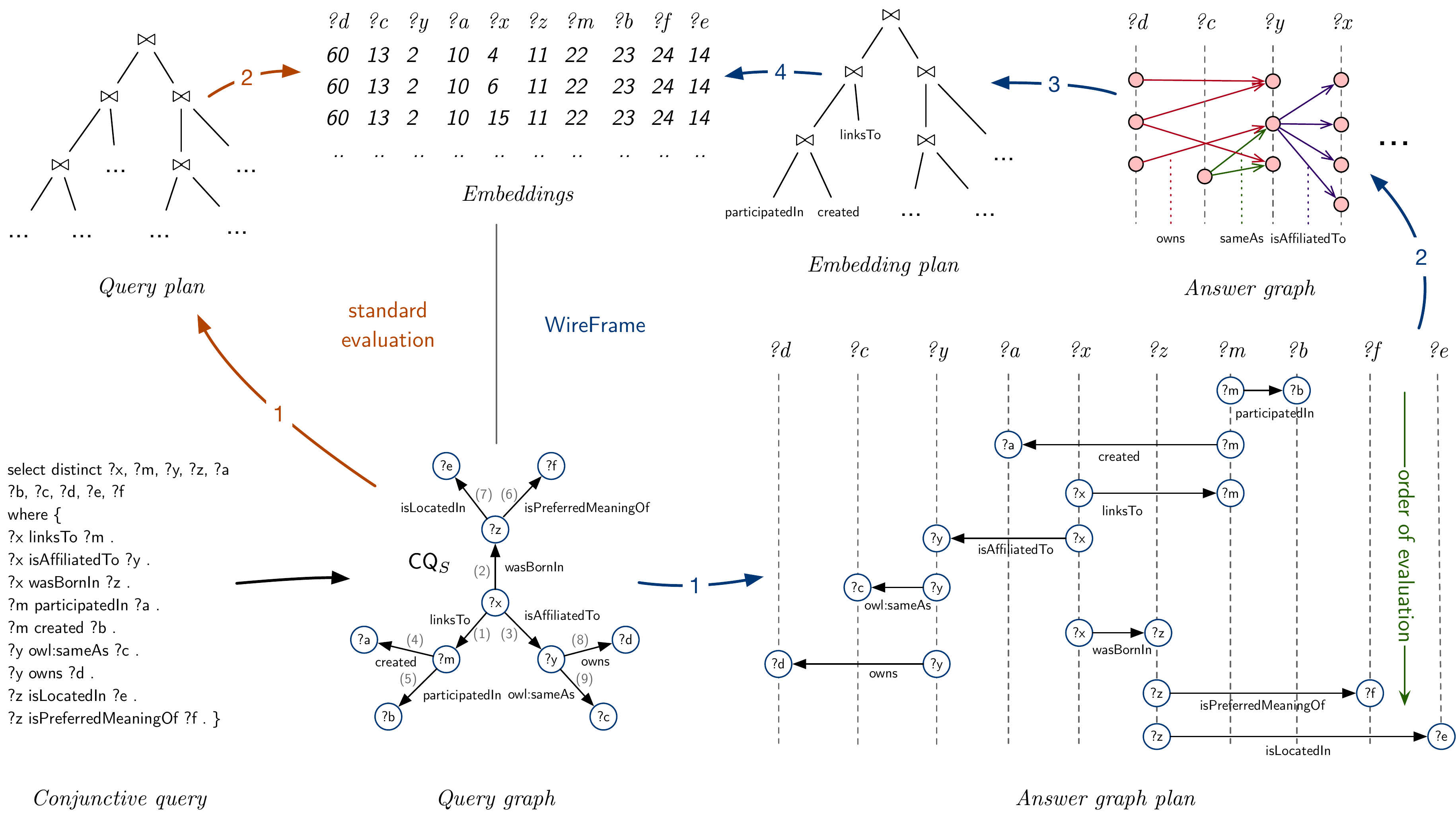}
	\caption{\wireframe:
	a two-phase cost-based optimizer for conjunctive queries.}
	\label{fig:wireframe}
\end{figure*}

Our evaluation model for \cq s
then becomes two phase:
\textit{answer-graph generation}
and
\textit{embedding generation}.

\smallskip

\noindent \textbf{Answer-graph generation.}
For each query edge of the plan, in turn,
our answer graph (\ag) is populated
with the matching labeled edges from \dg\
that meet the join constraints with the current state of the \ag.
Call this an \emph{edge-extension} step.
Then nodes in the \ag\ that failed to extend
are \emph{removed}.
This ``node burnback'' cascades.

Consider the \cq\ with query edges
\tuple{\var{?w}, \var{A}, \var{?x}},
\tuple{\var{?x}, \var{B}, \var{?y}},
and
\tuple{\var{?y}, \var{C}, \var{?z}}.
Fig.~\ref{fig:evaluation_model_example}
illustrates the interleaved
edge-extension and burn-back steps
over a sample data graph \dg.

\noindent \textbf{Embedding generation.}
The embedding tuples are then generated
over the answer graph
by joining the answer edges appropriately.
Given the \emph{ideal} answer graph
and an acyclic \cq,
the order in which we join
is immaterial.
No \(k\)-ary tuple is ever eliminated during a join
with a next query edge from the \iag.
This step is often quite fast,
given the \iag\ is small.
Evaluating this directly from the data graph \dg,
on the other hand%
---%
which is what other evaluation methods for \cq s do%
---%
can be exceedingly expensive.
(Fig.~\ref{fig:wireframe} illustrates,
comparing a standard evaluation with ours.)


\section{The Planners} \label{sec:framework}


\noindent
\textbf{I.~The Answer-graph Planner.}
\smallskip

\noindent
\textbf{Plan Cost.}
The \emph{edge walk} is our unit for estimating a plan's cost:
the retrieval of a matching edge from \dg.
To estimate the number of edge walks,
\emph{node} and \emph{edge} cardinality estimations are made
for each successive edge extension.
Note that the cost of node burnback is amortised:
every edge added that does not survive to the \iag\
is at some point removed.
\wireframe\ employs cardinality estimators
drawn from a catalog
consisting of
1-gram and 2-gram edge-label statistics
computed offline
\citep{%
    yakovets:2015:thesis,%
    yakovets:2016:QPE,%
    mannino1988statistical,%
    christodoulakis1989estimation
}. 


\noindent
\textbf{The Edgifier.}
A plan is a sequence of the \cq's query edges to be materialized.
We employ a bottom-up, dynamic-program\-ming algorithm
to construct the edge order based on cost estimation
(which relies upon the cardinality estimations).

\begin{figure}[t]
	\centering
	\includegraphics[width=\linewidth]{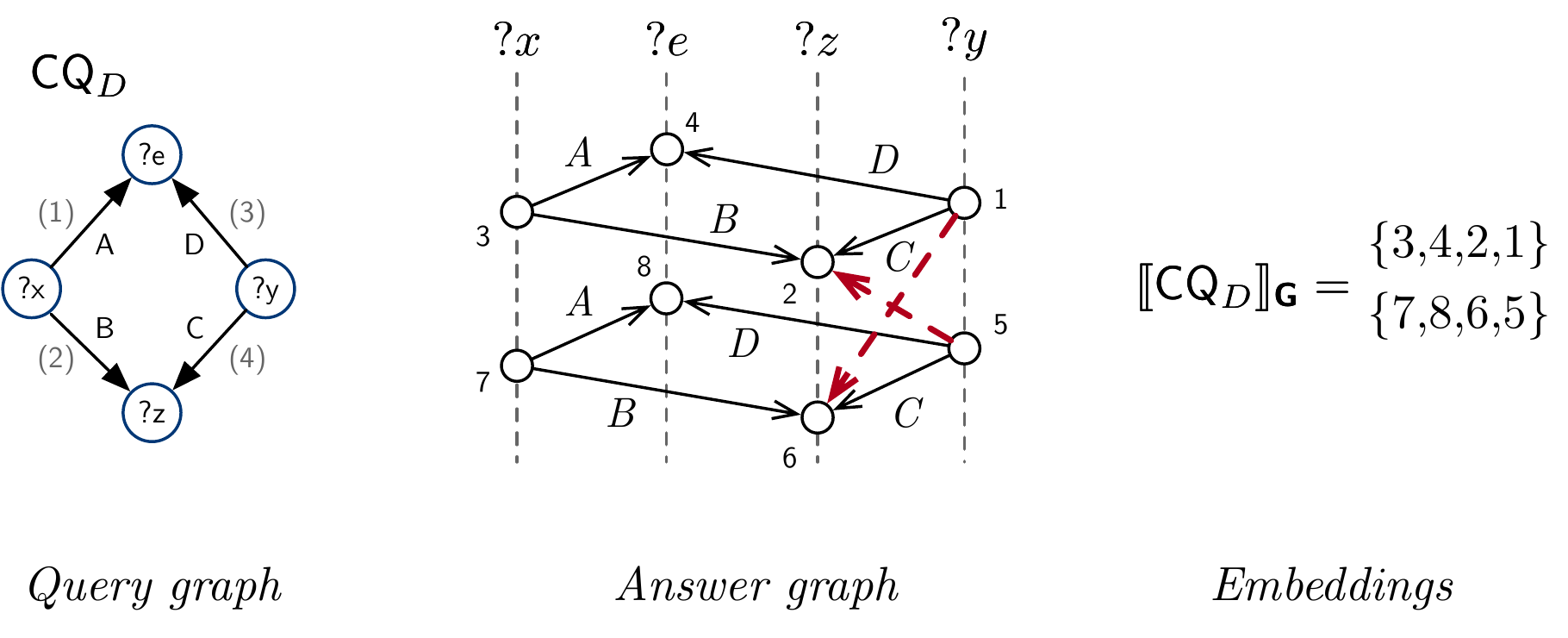}
	\caption{With only node burnback in a cyclic \textsf{CQ}.}
	\label{fig:counter_example_ideal_answer_graph}
\end{figure}

When the query graph of a \cq\ has cycles%
---%
a \emph{cyclic} query%
---%
there is an additional part to planning.
Node burn-back suffices to generate the ideal answer graph
for acyclic queries,
but not for cyclic.
The example
in Fig.~\ref{fig:counter_example_ideal_answer_graph}
illustrates why.
\emph{Spurious} edges%
---%
e.g., \tuple{1, 6}\
and \tuple{5, 2}%
---%
can remain that do not participate in any embedding.
To cull spurious edges requires an \emph{edge burnback} procedure
in addition to node burnback.
This requires the \cq 's cycles have been \emph{triangulated};
node \emph{triples} are materialized
in addition to the node \emph{doubles}
(the \ag\ edges)
during evaluation.
Triangulation is the choice
of which additional ``query edges'',
which we call \emph{chords},
to add.

\noindent
\textbf{The Triangulator.}
For cyclic \cq s,
in addition to the query-edge enumeration,
cycles in the query graph of length greater than three
are \emph{triangulated} by adding \emph{chord} edges.
We employ a bottom-up dynamic programming algorithm
to generate a bushy plan
that dictates the order and choice
of chord bisection of cycles (down to triangles).
During evaluation,
a chord is maintained as the intersection of
the materialized joins
of the opposite two edges for each triangle in which it participates.

The chordified plan when executed with node burnback guarantees
that the node sets will always be minimal.
A correct answer graph, \ag, will be found,
but it is no longer guaranteed to be ideal.
This is because spurious edges may remain
in the \ag,
as demonstrated
in Fig.~\ref{fig:counter_example_ideal_answer_graph}.
The embeddings can, of course, be found
from the non-ideal \ag\ established.

\noindent
\textbf{Edge Burnback.}
With the addition of an \emph{edge burn-back} mechanism at runtime,
we can guarantee again
that we find the ideal \ag\ (\iag).
This works by checking the chords' materializations
to chase what needs to be removed on cascade.
This ensures that spurious edges are removed.
The additional overhead of edge burn\-back
must be balanced off
against the benefit of obtaining the \iag\
versus a larger, non-ideal \ag.
This is work in progress.
In our experiments,
our evaluation over cyclic \cq s
is without edge burnback.


\begin{table*}[]
\centering

\resizebox{\textwidth}{!}{{\sffamily
\begin{tabular}{rl|rrrrr|rr}
\toprule
$\cq_S$ & \textbf{Snowflake-shaped Queries} (1/2/3/4/5/6/7/8/9)                                                                         & PG & WF  & VT  & MD & NJ  & |iAG|   & |Embeddings|    \\
\midrule
1 & diedIn/influences/actedIn/owns/wasCreatedOnDate/actedIn/created/hasDuration/ wasCreatedOnDate      & 51 & \textbf{16}  & *   & *  & * & 1660   & 2931986  \\
2 & hasChild/influences/actedIn/actedIn/wasBornIn/created/actedIn/hasDuration/wasCreatedOnDate        & 88 & \textbf{5}   & 151 & *  & * & 993    & 2847184  \\
3 & isCitizenOf/influences/actedIn/exports/wasCreatedOnDate/actedIn/created/hasDuration/wasCreatedOnDate & 69 & \textbf{12} & * & * & * & 1140 & 2670339 \\
4 & isMarriedTo/influences/actedIn/actedIn/wasBornOnDate/created/actedIn/hasDuration/wasCreatedOnDate & 78 & \textbf{8}   & *   & *  & * & 3317   & 2569017  \\
5 & isMarriedTo/diedIn/actedIn/actedIn/wasBornIn/owns/wasCreatedOnDate/hasDuration/wasCreatedOnDate   & 42 & \textbf{12}  & *   & *  & * & 10761  & 1306406  \\
\midrule \midrule
$\cq_D$ & \textbf{Diamond-shaped Queries} (1/2/3/4)                                                                                            & PG & WF  & VT  & MD & NJ  & |AG|   & |Embeddings|    \\ 
\midrule
6 & livesIn/isCitizenOf/isLocatedIn/linksTo                                                           & *  & \textbf{103} & *   & *  & *   & 833355 & 58785214 \\
7 & livesIn/isCitizenOf/linksTo/happenedIn                                                            & *  & 118 & \textbf{38}  & *  & 127 & 22555  & 100160   \\
8 & diedIn/linksTo/wasBornIn/graduatedFrom                                                            & *  & \textbf{20}  & 110 & *  & 213 & 68720  & 106214   \\
9 & diedIn/linksTo/wasBornIn/isLeaderOf                                                               & *  & \textbf{18}  & 22  & *  & 139 & 87459  & 22216    \\
10 & diedIn/linksTo/wasBornIn/hasWonPrize                                                              & *  & \textbf{53}  & 126 & *  & *   & 52975  & 99891    \\  
\bottomrule 
\end{tabular}
}}
\caption{Query execution time (sec) in different systems (* denotes terminating the query after 300 seconds).}
\label{tbl:result}
\end{table*}

\smallskip
\noindent
\textbf{II. The Embedding Planner.}

\noindent
\textbf{Plan Cost.}
When generating the embeddings for an acyclic \cq\
from its \iag,
the order in which we join (connected) answer edges is
immaterial.
As the \(k\)-ary tuples are extended,
no intermediate results are ever lost.
Thus,
for this,
no planning is required.


\noindent
\textbf{The Defactorizer.}
On the other hand,
when the \cq\ is cyclic or the \ag\ provided is non-ideal,
intermediate results can be lost.
The join order then matters.
We call this process \emph{defactorization}.
Alternative plans for embedding materialization 
are synonymous with choosing this join order.
It is possible to do this again via a cost-based approach
via a bottom-up, dynamic programming algorithm,
using our catalog statistics.


\section{Experiments}\label{results}


\noindent \textbf{Prototype.}
We have implemented a prototype, \wireframe, which runs on top of \postgresql,
a popular relation database system. \wireframe\ implements the two phases described in Section \ref{sec:framework}, each with a separate planner and evaluator. The planner for the first phase outputs an optimal left-deep tree plan that indicates the execution order of the query edges. The evaluator then takes the tree plan to evaluate the query edges in sequence. For the second phase, we presently use a greedy approach to generate a tree plan based on the available statistics from the answer graph phase. The node burnback procedure is implemented via procedural SQL.


\noindent \textbf{Environment.}
For evaluating \wireframe's performance,
we use the YAGO2s dataset \citep{Yago2s}.
With a select set of five acyclic and five cyclic \cq s,
we compare query execution times against
    \postgresql\ v11.0 (PG),
    \virtuoso\ v6.01 (VT),
    \monetdb\ v11.31 (MD),
        and
    \neofourj\ v.3.5 (NJ).
All experiments were conducted on a server running \ubuntu\ 18.04 LTS
with two Intel Xeon X5670 processors and 192GB of RAM.
After preprocessing the dataset
(contains 242M triples with 104 distinct predicates),
we imported it to each of the systems.

For the queries,
we implemented a query miner that generates queries
over a dataset
using query templates
(with placeholders for edge labels).
The query miner then generates valid, non-empty queries.
For our experiments,
we use two templates, $CQ_{S}$ and $CQ_{D}$,
as shown
in Figures \ref{fig:wireframe} and \ref{fig:counter_example_ideal_answer_graph},
respectively.
With these two templates,
we mined 218,014 snowflake-shaped queries
and 18,743 diamond-shaped queries.
For our preliminary experimental study,
we chose five queries of each shape
which could be attributed to real-life use cases.

For \postgresql\ and \monetdb,
the dataset was imported as a triple store,
with indexes on the string dictionary,
and six composite indexes
over the permutations
of \emph{subject}, \emph{predicate}, and \emph{object}.
We set the size of the memory pool to eight GB
for all of the systems,
except for \monetdb\
(which sets its own resource allocations based on the server).
We repeat execution of each query five times,
taking the average of the last four runs (\emph{i.e.,} warm cache),
as reported in Table \ref{tbl:result}.
The execution time is the time spent
to retrieve all the result tuples for a query.  


\noindent \textbf{Results.}
One can observe
from Table \ref{tbl:result}
that the size of the answer graph is exceedingly smaller
than the number of embeddings.
For instance,
for the second snowflake-shaped query,
the \ag\ is 2,867 times smaller
than the number of embeddings.
It is no surprise, therefore,
that \wireframe\ (WF) achieves good performance;
it avoids the redundant edge-walks that arise from many-many joins.
While the second snowflake-shaped query took 88 seconds on \postgresql,
it only took five seconds on  \wireframe.
The answer-graph approach requires
a much smaller memory footprint,
which can be beneficial for traditional database systems
that heavily use secondary storage.
The approach also competes well
against main-memory intense systems
such as \neofourj\ and \virtuoso.
For the cyclic, diamond-shaped queries,
employing only node burnback does not guarantee the ideal answer graph,
as discussed above.
We have found that the resulting \ag s can be significantly larger
than the ideal,
sometimes close to the number of embeddings.
For this reason,
\wireframe\ was slower for some of the cyclic queries, notably 1 and 2.
Even so,
its performance over cyclic queries is quite good.
With further plan- and run-time optimization with edge burnback,
we believe that
the performance will be stellar.


\section{Conclusions}  

We have clear objectives for our next steps.
\emph{First},
one has a richer plan space
when considering bushy plans
for both our first and second phases.
The challenge is to devise a suitable cost model
for searching the bushy-plan space via dynamic programming.
\emph{Second},
when the size of an answer graph is distant from the ideal,
generating the embeddings can be costly.
Triangulation promises to reduce this significantly.
This requires investigating the trade-offs
between the added cost for maintaining the triangle materializations
and the reduced cost from generating the embeddings
from the significantly smaller ideal \ag.
\emph{Lastly},
we are to explore further optimizations within this space.
Large graphs are meant to be queried.



%
%
%

\balance
\bibliographystyle{abbrvurl}
\bibliography{paper}

\end{document}